\documentclass[twocolumn,showpacs,superscriptaddress,floatfix,prb]{revtex4-1}
\usepackage{afterpage}
\usepackage[caption=false]{subfig}
\usepackage{multirow}
\usepackage{graphicx,times}  
\usepackage{soul}
\usepackage{natbib}
\usepackage{amssymb,amsmath}
\usepackage{amsfonts}
\usepackage{amsbsy}
\usepackage{comment}

\usepackage{color}
\usepackage{float}

\usepackage[colorlinks=true, urlcolor=blue, citecolor=blue, linkcolor=blue]{hyperref}
\usepackage{color}
\usepackage{tikz}
\usepackage{mathtools} 
\usepackage{braket}
\usetikzlibrary{positioning,arrows}
\usetikzlibrary{decorations.pathmorphing}
\usetikzlibrary{decorations.markings}
\definecolor{myblue}{RGB}{56,94,141}
\renewcommand{\vec}[1]{\mbox{\boldmath$\mathrm{#1}$}}

\renewcommand{\vec}[1]{\mbox{\boldmath$\mathrm{#1}$}}
\newcommand{\be}{\begin{equation}}
\newcommand{\ee}{\end{equation}}
\newcommand{\ben}{\begin{eqnarray}}
\newcommand{\een}{\end{eqnarray}}
\begin{document}


\title{Plasmonic skyrmion quantum thermodynamics}
\author{Vipin Vijayan$^1$, L. Chotorlishvili$^2$, A. Ernst$^{3,4}$, Mikhail I. Katsnelson$^{5}$, S. Parkin$^{3}$, Sunil K. Mishra$^1$}
\address{$^1$ Department of Physics, Indian Institute of Technology (Banaras Hindu University), Varanasi - 221005, India\\
$^2$Department of Physics and Medical Engineering, Rzesz\'ow University of Technology, 35-959 Rzesz\'ow, Poland\\
$^3$ Max Planck Institute of Microstructure Physics, Weinberg 2, D-06120 Halle, Germany\\
$^4$ Institute for Theoretical Physics, Johannes Kepler University, Altenberger Strasse 69, 4040 Linz, Austria\\
$^5$ Radboud University, Institute for Molecules and Materials, Heyendaalseweg 135, 6525 AJ Nijmegen, the Netherlands}

\date{\today}

\begin{abstract}
The primary obstacle in the field of quantum thermodynamics revolves around the development and practical implementation of quantum heat engines operating at the nanoscale. One of the key challenges associated with quantum working bodies is the occurrence of ``quantum friction," which refers to irreversible wasted work resulting from quantum inter-level transitions. Consequently, the construction of a reversible quantum cycle necessitates the utilization of adiabatic shortcuts. However, the experimental realization of such shortcuts for realistic quantum substances is exceedingly complex and often unattainable. In this study, we propose a quantum heat engine that capitalizes on the plasmonic skyrmion lattice. Through rigorous analysis, we demonstrate that the quantum skyrmion substance, owing to its topological protection, exhibits zero irreversible work. Consequently, our engine operates without the need for adiabatic shortcuts.  We checked by numerical calculations and observed that when the
system is in the quantum skyrmion phase, the propagated states differ from the initial states only by the geometricl and dynamical phases. The adiabacit evoluation leads to the zero transition matrix elements
and zero irreversible work. By employing plasmonic mods and an electric field, we drive the quantum cycle. The fundamental building blocks for constructing the quantum working body are individual skyrmions within the plasmonic lattice. As a result, one can precisely control the output power of the engine and the thermodynamic work accomplished by manipulating the number of quantum skyrmions present.

\end{abstract}

\maketitle
Quantum thermodynamics investigates thermodynamic processes at the quantum level \cite{PhysRevResearch.2.043247, PhysRevE.101.020201, PhysRevLett.124.040602, PhysRevResearch.2.023377, PhysRevE.102.040103} and establishes a link between thermodynamics and quantum theory. The practical aspect of quantum thermodynamics is constructing quantum heat engines and study of quantum thermodynamic processes \cite{PhysRevResearch.2.023120, kosloff2017quantum, PhysRevResearch.2.032062, PhysRevLett.125.166802, PhysRevLett.124.110604, RevModPhys.91.045001, abah2017energy, PhysRevE.98.032121, azimi2014quantum, chotorlishvili2016superadiabatic, chotorlishvili2011thermal}.  Quantum heat engines can be incorporated into nanodevices and perform work on the nano level. The fundamental aspect of quantum thermodynamics is the study of the role of quantumness in thermodynamic processes, which is highly important for the foundation of quantum statistical mechanics. In the present work, we analyze both fundamental and practical aspects. 
For a comprehensive understanding of various aspects of quantum thermodynamics and quantum heat engines, the recent literature provides valuable insights \cite{PhysRevE.94.062109, PhysRevE.99.022110, Stefanatos2018, delCampo2019, del2018friction,
PhysRevE.98.022107, PhysRevE.100.032144, PhysRevE.102.030102, PhysRevB.100.085405, chattopadhyay2019relativistic, pena2019magnetic,
altintas2019comparison, pena2017magnetic,Sotnikov2023, PhysRevA.101.022113, PhysRevE.101.032113, PhysRevResearch.2.023145, arXiv:2003.05788}.
Experimental realizations of quantum heat engines have already been achieved \cite{rossnagel2016single,PhysRevLett.112.076803,Goldwater2019,
PhysRevLett.122.110601}. The critical effect of quantumness is the effect of quantum friction. Swift driving of a quantum system leads to the quantum multiple inter-level transitions. As a result, a substantial amount of work is wasted in the irreversible work. The irreversible work can be reduced by slowing down the driving speed of the cycle. However, this reduces the output power of the quantum engine as well. The elegant method is constructing the adiabatic shortcuts, an extra driving term compensating irreversible losses of quantum origin \cite{campo2014more,demirplak2003adiabatic,berry2009transitionless}. Unfortunately, adiabatic shortcuts are hard to construct for experimentally feasible physical systems. In the present work, we propose a novel solution to the problem based on the effect of topological protection. We analyze quantum thermodynamic processes and consider quantum skyrmion as a working body for the quantum Otto cycle. We will prove that due to the topological projection, when the system is in the quantum skyrmion phase, propagated in the time quantum states differ from the initial states only by dynamical and geometrical phases (calculated and presented in supplimentary materials). Consequently, nondiagonal transition matrix elements are zero leading to zero irreversible work. The evolution of the system in the quantum skyrmion phase is adiabatic.  Before considering thermodynamic aspects, for the interest of a broad audience, we briefly review the field of quantum skyrmionics.

\begin{figure}
    \centering
    \includegraphics[width=\columnwidth]{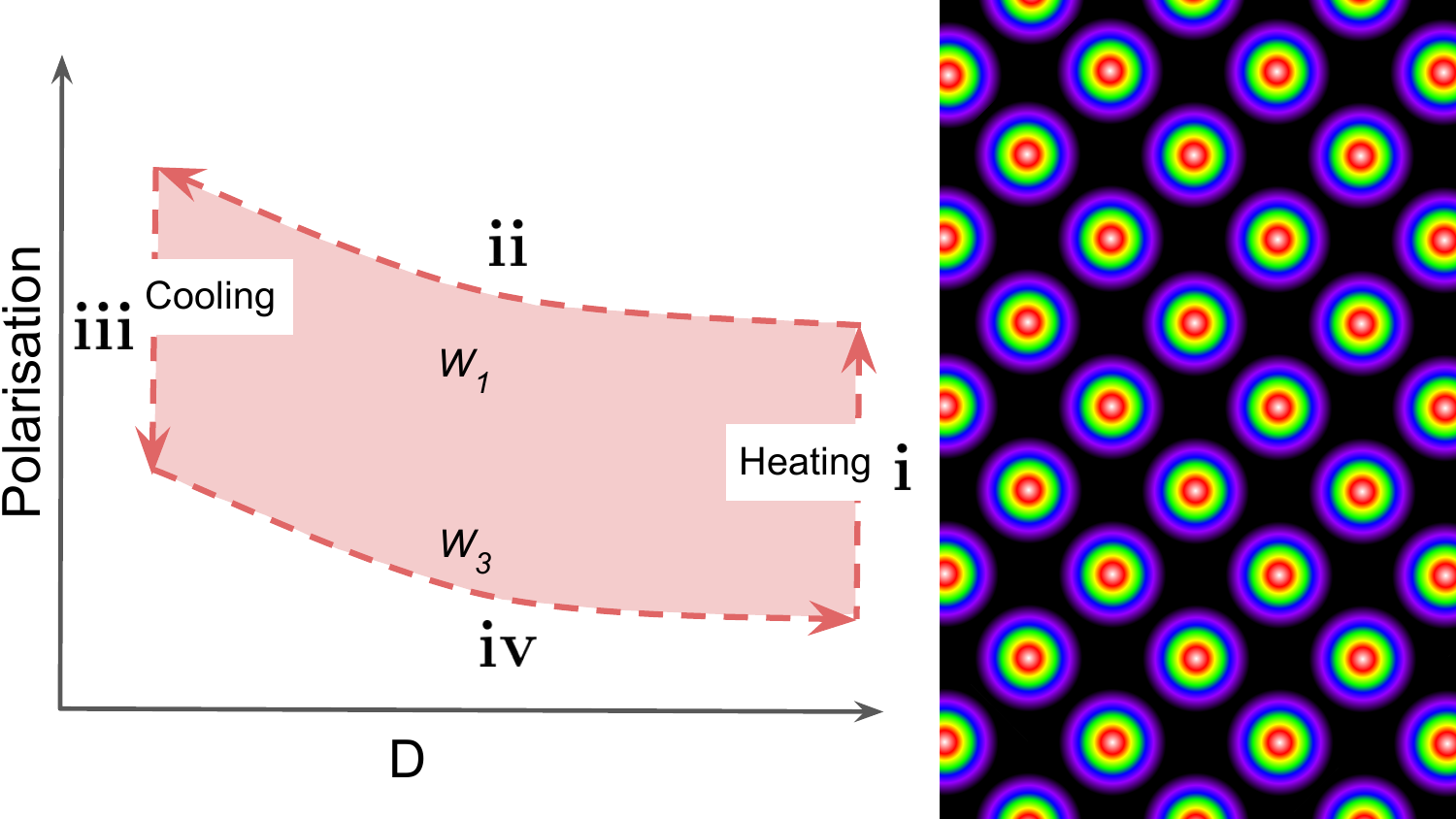}
    \caption{Schematics of the heating/cooling cycle (left) and the plasmonic skyrmion lattice (right).}
    \label{fig:schematic}
\end{figure}

Particle-like topological solitons, particularly skyrmions, exhibit great promise for spintronics applications. Skyrmions were initially discovered in studies of fundamental field theory \cite{piette1995multisolitons,rajaraman1982solitons,skyrme1994non,belavin1975metastable} and subsequently observed in magnetic systems \cite{barton2020magnetic,schroers1995bogomol,seki2012observation,wilson2014chiral,
schutte2014magnon,PhysRevLett.129.126101,PhysRevB.106.104424}. The underlying physical mechanism responsible for the formation of skyrmion magnetic textures involves the Dzyaloshinskii–Moriya interaction (DMI) or spin frustration, arising from the competition between nearest-neighbor ferromagnetic and next-nearest-neighbor antiferromagnetic exchange interactions. In the field of condensed matter physics, the term "skyrmionics" has emerged as a prominent topic of research \cite{white2014electric,derras2018quantum,haldar2018first,leonov2015multiply,psaroudaki2017quantum,van2013magnetic,rohart2016path,samoilenka2017gauged,battye2013isospinning,jennings2014broken,tsesses2018optical}. Classical magnetic skyrmions are formed by a collective arrangement of numerous spins. However, recent attention has shifted towards quantum skyrmions \cite{PhysRevB.107.L100419,OchoaTserkovnyak,PhysRevB.98.024423,stepanov2019heisenberg,PhysRevX.9.041063,PhysRevResearch.4.023111,PhysRevB.103.L060404,PhysRevB.107.L100419}.
In a recent work \cite{PhysRevLett.125.227201}, a new concept of matter called a "plasmonic skyrmion lattice" was suggested. This lattice is formed within an optical system through the interference patterns of surface plasmon polaritons generated by coherent or incoherent laser sources. The skyrmions, which are topologically nontrivial magnetic structures, become confined to the nodal points of this optical lattice. The underlying mechanism responsible for this confinement is the magnetoelectric (ME) effect. The emergence of a ferroelectric polarization \cite{PhysRevLett.95.057205}, denoted as $\mathbf{P}=g_{ME}\vec{e}_{ij}\times(\hat{S}_i\times\hat{S}_j)$, couples with the electric field of the plasmons, resulting in an effective Dzyaloshinskii-Moriya interaction term characterized by $D=g_{ME}E$. Here, $\vec{e}_{ij}$ represents the unit vector connecting neighboring spins $\mathbf{\hat S_i}$ and $\mathbf{\hat S_j}$, and $g_{ME}$ denotes the constant associated with the magnetoelectric coupling. It is worth noting that experimental observations of skyrmions in a single-phase multiferroic material, Cu$_{2}$OSeO$_{3}$, have been reported \cite{seki2012observation}. For insights into the quantum aspects, the work \cite{janson2014quantum} provides relevant information.
As it becomes evident from the definition, the noncolinear magnetic spin texture $(\mathbf{\hat S_i}\times\mathbf{\hat S_j})$ is a crucial aspect of the magnetoelectric (ME) effect. Consequently, when an external electric field is applied to a magnetic layer, it selectively couples with skyrmions rather than the majority of the surface layer that exhibits ferromagnetic spin order. This observation opens up new possibilities in the field of quantum thermodynamics. In this paper, we propose a novel type of quantum heat engine called the ``\textit{plasmonic skyrmion quantum heat engine}." Additionally, previous research \cite{PhysRevLett.125.227201} has shown that the distances between skyrmions in a plasmonic lattice can be adjusted to the range of 250-300 nm. Consequently, noninteracting skyrmions can be treated as individual modules within the quantum engine. By employing an arbitrary number of skyrmions, the working substance of the plasmonic skyrmion heat engine can be tailored to achieve the desired work and output power. A typical quantum heat cycle consists of four stages, during which the quantum working body is connected to cold and hot heat baths and externally driven to produce thermodynamic work (Fig. \ref{fig:schematic}). In the subsequent sections, we demonstrate that the topological protection of quantum skyrmions nearly eliminates irreversible work. Consequently, the plasmonic quantum skyrmion heat engine does not require adiabatic shortcuts, offering a substantial advantage over other models.

\textbf{\textit{Model}}: Before delving into the thermodynamic properties of a quantum skyrmion, it is essential to define the specific model under consideration, as discussed in recent research \cite{PhysRevResearch.4.023111}. 
\begin{eqnarray}
\label{pia_Hamiltonian}
 \hat H=&-J\sum\limits_{\langle i,j\rangle}(\hat S_i^x\hat S_j^x + \hat S_i^y\hat S_j^y)-\Delta \sum\limits_{\langle i,j\rangle}\hat S_i^z\hat S_j^z\nonumber\\
 &-D\sum\limits_{\langle i,j\rangle}(\mathbf{e_{ij}}\times \hat{z}).\left(\mathbf{\hat S}_i\times\mathbf{\hat S}_j\right).
\end{eqnarray}
In our model, we consider a ferromagnetic exchange constant $J>0$, which serves as the energy scale for the problem and is set to $J=1$ for convenience. The axial Heisenberg anisotropy is characterized by a positive parameter $\Delta$, while the strength of the Dzyaloshinskii-Moriya interaction is denoted by the constant $D$. 
The lattice consists of $n \times n$ sites, with each site hosting a spin-1/2 described by $\mathbf{\hat{S}_i}=\frac{\hbar}{2}\mathbf{\hat\sigma}$. 
Importantly, the plasmonic modes in the system only couple with the noncollinear magnetic textures of the skyrmions. By controlling the strength of the DM interaction $D$ through the plasmonic mode $E$ \cite{PhysRevLett.125.227201}, we can tune the parameter $D=g_{ME}E$ and define the region of the skyrmion phase.

For verifying the existence and stability of the quantum skyrmion, we exploit two quantities \cite{Gauyacq,berg}:  the magnitude of the winding parameter $Q$, and the topological index $C$  defined as follows:
\begin{equation}
\begin{rcases}
  Q \\
   C
 \end{rcases} = \frac{1}{2\pi} \sum \limits_{\sigma} \tan^{-1}\left(\frac{ \mathbf{n}_i(\mathbf{n}_j \times \mathbf{n}_k)}
 {1+ \mathbf{n}_i \mathbf{n}_j + (\mathbf{n}_i \mathbf{n}_k)(\mathbf{n}_k\mathbf{n}_j)}\right).
\end{equation}
In our system, the summation is performed over all elementary triangles formed by nearest-neighbor lattice sites $i$, $j$, $k$, including the classical ferromagnetic boundary sites, with no overlapping triangles. The stability of the skyrmion is characterized by the winding parameter $Q$, which is calculated as $\mathbf{n_i} = 2 \langle \mathbf{S}_i \rangle /\hbar$, where $\langle \mathbf{S}_i \rangle = (\langle S_i^x \rangle, \langle S_i^y \rangle, \langle S_i^z \rangle)$ represents the spin expectation value. The topological index $C$ is defined as $\mathbf{n_i} =  \langle \mathbf{S}_i \rangle / 
 |\langle \mathbf{S}_i \rangle|$, and it takes the values $C = \pm 1$ for quantum skyrmions (antiskyrmions). 
 
 It is worth mentioning that topological entanglement entropy (TEE) is a useful quantity for identifying phases and assessing topological protection, particularly in larger system sizes \cite{vijayan1}. However, in our case, since we are working with small system size and employing open boundary conditions, TEE does not provide significant insights.

 \begin{figure}
     \centering
         \includegraphics[scale=.5]{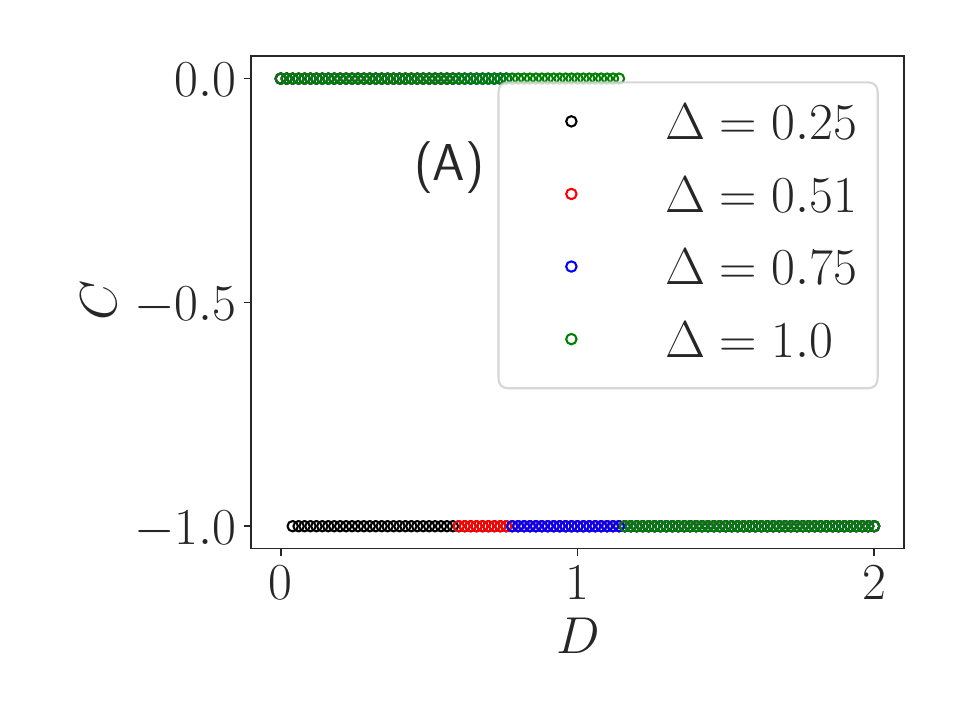} \vskip -1cm 
         \includegraphics[scale=.5]{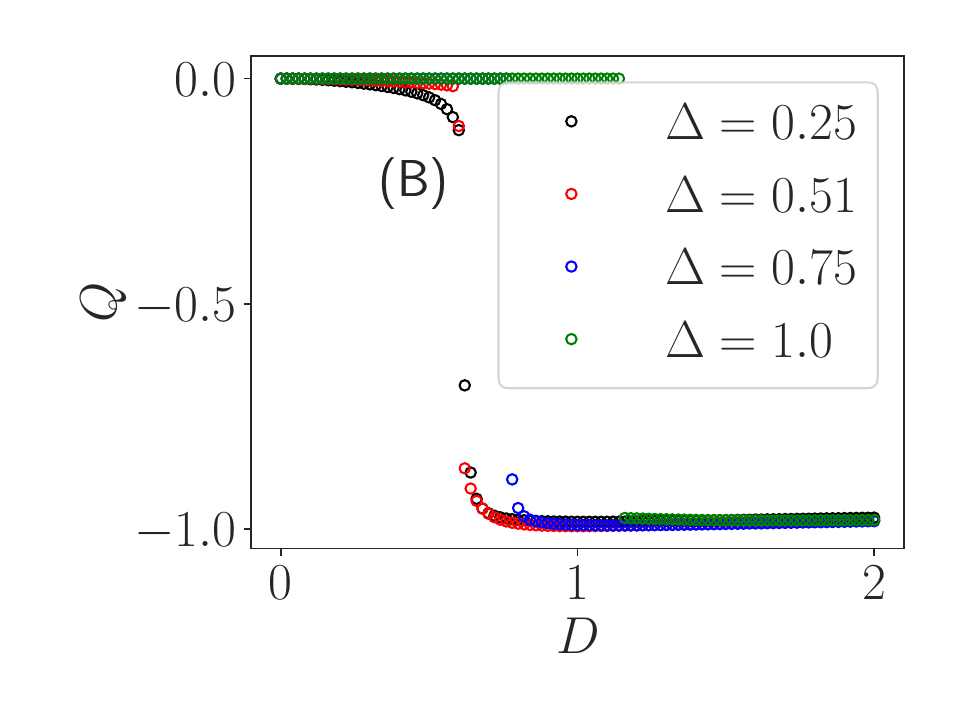} 
     \caption{A) Topological Index, $C$ v/s $D$ magnitude for various anisotropy constant values. B) Winding parameter, $Q$ v/s $D$ magnitude for various anisotropy constant values.}
     \label{fig:exact results}
 \end{figure}

As shown in Fig:~\ref{fig:exact results}(A), the system exhibits either a ferromagnetic state with $C=0$ or an antiskyrmion state with $C=-1$, depending on the values of the anisotropy parameter $\Delta$ and the plasmonic field $D$. From Fig:~\ref{fig:exact results}(B) we have shown the value of $Q$ whose magnitude aligns with the stability of the corresponding skyrmion state. For any value of $\Delta$, we see a significant increase in $Q$; hence, stability is observed around $D=0.8J$. We also notice that smaller $\Delta$ preferably forms stable skyrmions even at small values of $D$.  In the following analysis, we will focus on the antiskyrmion state and investigate the first thermodynamic quantity of interest, namely non-equilibrium work. To describe the system's state, we define the thermal density matrix as $\hat{\rho}_0 = Z^{-1}\exp(-\beta E_n)\ket{\varphi}\bra{\varphi}_n$, where $\beta$ is the inverse temperature, $Z=\sum_ne^{-\beta E_n}$ is the partition function, $\ket{\varphi}_n$ represents the eigenfunctions of the system described by Eq.(\ref{pia_Hamiltonian}), including low-lying skyrmionic states \cite{PhysRevB.107.L100419}, and $E_n$ are the corresponding eigenvalues. We drive the system by manipulating the plasmonic field $D(t)$ in time, specifically within the antiskyrmion region. The skyrmion, acting as the working substance, produces thermodynamic work, which can be quantified as 
\begin{equation}
\langle W\rangle=\langle W_{\text{ir}}\rangle+\Delta F, 
\label{work done}
\end{equation}
where $\langle W\rangle$ represents the total work produced by the skyrmion quantum heat engine, $\Delta F=-\frac{1}{\beta}\ln(Z_f/Z_0)$ corresponds to the change in free energy during the stroke, and $\langle W_{\text{irr}}\rangle$ denotes the irreversible work. Irreversible work arises due to quantum inter-level transitions occurring in finite-time processes. It represents the dissipated or wasted work and can be quantified by the deficit between the produced work and the change in the system's free energy \cite{delCampo2019,del2018friction,PhysRevResearch.2.032062,abah2017energy,chotorlishvili2016superadiabatic}.

The quantum Otto cycle consists of the following four strokes, as illustrated in Fig.~\ref{fig:schematic}:

\textbf{i)} Thermalization at high temperature: At the beginning of the cycle, the magnetic texture (skyrmion) is in contact with a high-temperature bath at temperature $T_H$. By applying an electric field $E_0$, the magnetoelectric coupling results in a change in the Dzyaloshinskii-Moriya interaction parameter, $D_0 = g_{ME}E_0$, at time $t=0$. During this stroke, the skyrmion absorbs heat $Q_{in}$ from the high-temperature bath.

\textbf{ii)} Adiabatic work extraction: The electric field is changed from $E_0$ to $E_1$ during this stroke, where $E_1>E_0$. This change in the electric field corresponds to a change in the Dzyaloshinskii-Moriya interaction parameter, $D_1 = g_{ME}E_1$, at time $t=\tau$. The duration of this stroke is denoted by $\tau$. The system performs adiabatic work, denoted as $W_2$, during this process. The work done is given by the expression $W_2 = \sum_n \left(\mathcal{E}_n(\tau) - \mathcal{E}_n(0)\right) P_{n,2}(\beta_H, \mathcal{E}(0))$. 
Here, $\mathcal{E}_n$ represents the energy eigenvalues, and $\beta_H=1/T_H$ is the inverse temperature. The probability $P_{n,2}(\beta_H, \mathcal{E}(0))$ is given by $P_{n,2}(\beta_H, \mathcal{E}(0)) = {Z(\beta_H, \mathcal{E}(0))e^{-\beta_H \mathcal{E}_n(0)}}$,
where $Z(\beta_H, \mathcal{E}(0)) =\left(\sum_n e^{-\beta_H \mathcal{E}_n(0)}\right)^{-1}$ is the partition function at the initial energy.

\textbf{iii)} Cooling through a cold bath: In this stroke, the system is brought into contact with a cold bath at temperature $T_L$. The cooling process takes place, allowing the system to release heat to the cold bath.

\textbf{iv)} Return to the initial state: The electric field/DMI is steered back from $E_1/D_1$ to the initial values $E_0/D_0$ during this stroke. The adiabatic work done on the working substance is denoted as $W_4$. It is given by: $W_4 = \sum_n \left(\mathcal{E}_n(0) - \mathcal{E}_n(\tau)\right) P_{n,4}(\beta_L, \mathcal{E}(\tau))$, where $\beta_L=1/T_L$ is the inverse temperature of the cold bath. The probability $P_{n,4}(\beta_L, \mathcal{E}(\tau))$ is given by: $P_{n,4}(\beta_L, \mathcal{E}(\tau)) = {Z(\beta_L, \mathcal{E}(\tau))e^{-\beta_L \mathcal{E}_n(\tau)}}$. Here, $Z(\beta_L, \mathcal{E}(\tau)) =\left(\sum_n e^{-\beta_L \mathcal{E}_n(0)}\right)^{-1}$ is the partition function at the final energy.

In summary, the quantum Otto cycle involves isothermalization, adiabatic work extraction, cooling, and return to the initial state. The work done and heat exchanged is determined by the energy eigenvalues, partition functions, and temperatures of the thermal baths involved.

In practice, the irreversible work can be calculated using the Kullback-Leibler distance between two density matrices $\hat\varrho_t$ and $\hat\varrho_t^{eq}$ which is given by \cite{Parrondo}: 
\begin{eqnarray} 
\label{Kullback-Leibler divergence}
&& \langle W_{ir}\rangle=\frac{1}{\beta}S\left(\hat\varrho_t\Vert\hat\varrho_t^{eq}\right),\nonumber\\
&& S\left(\hat\varrho_t\Vert\hat\varrho_t^{eq}\right)=\sum_n p_n^0\ln p_n^0-\sum_{n,m}p_n^0p_{m,n}^\tau\ln p_n^\tau .
\end{eqnarray}
Here $\hat\varrho_t$ and $\hat\varrho_t^{eq}=e^{-\beta\hat H(t)}/Tr\left[e^{-\beta\hat  H(t)}\right]$ are non-equilibrium and equilibrium density matrices, $p_{m,n}^\tau=\vert\langle \mathcal{E}_n(\tau)\vert \hat U(\tau)\vert \mathcal{E}_m(0)\rangle\vert^2$ is the inter-level transition probability, $\hat U(\tau)=e^{-i/\hbar\int^\tau_0 H(t)dt}$ is the time evolution operator, $p_n^0=e^{-\beta \mathcal{E}_n(0)}Z(\beta,\mathcal{E}(0))$ and
$p_n^\tau=\sum_m p_{m,n}^\tau p_m^0$ are level populations of the propagated state. From Eq.(\ref{Kullback-Leibler divergence}) is is easy to see that the irreversible work is zero if $p_{m,n}^\tau=\delta_{nm}$. This is also proved by numerical calculations done for the quantum skyrmion phase, and presented in the supplementary materials.

\begin{figure}
    \centering
        \includegraphics[scale=0.5]{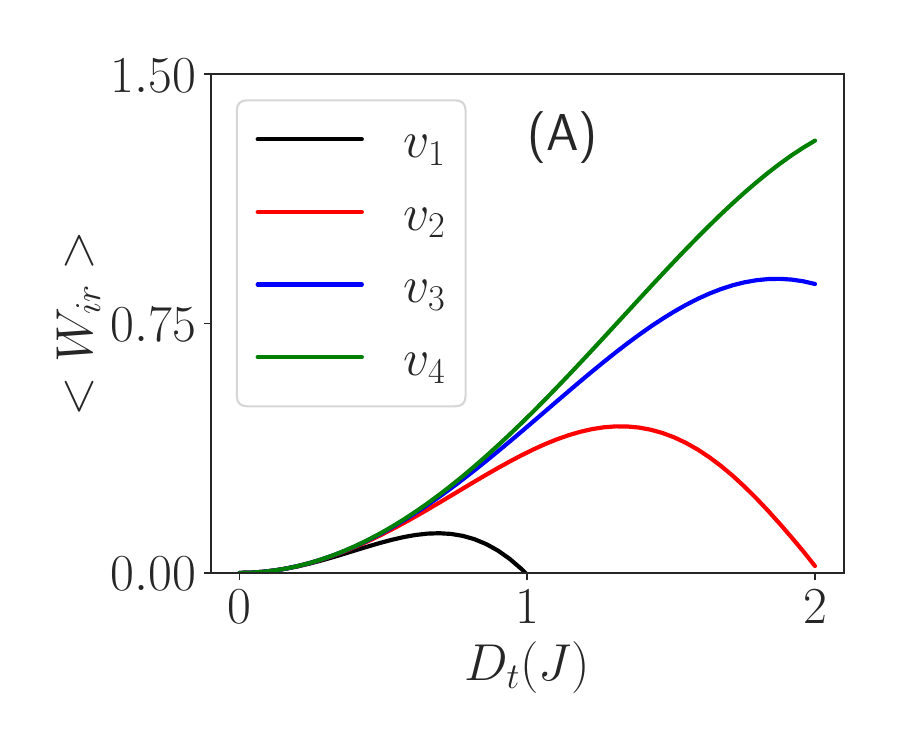}  \vskip -1cm
        \includegraphics[scale=0.537]{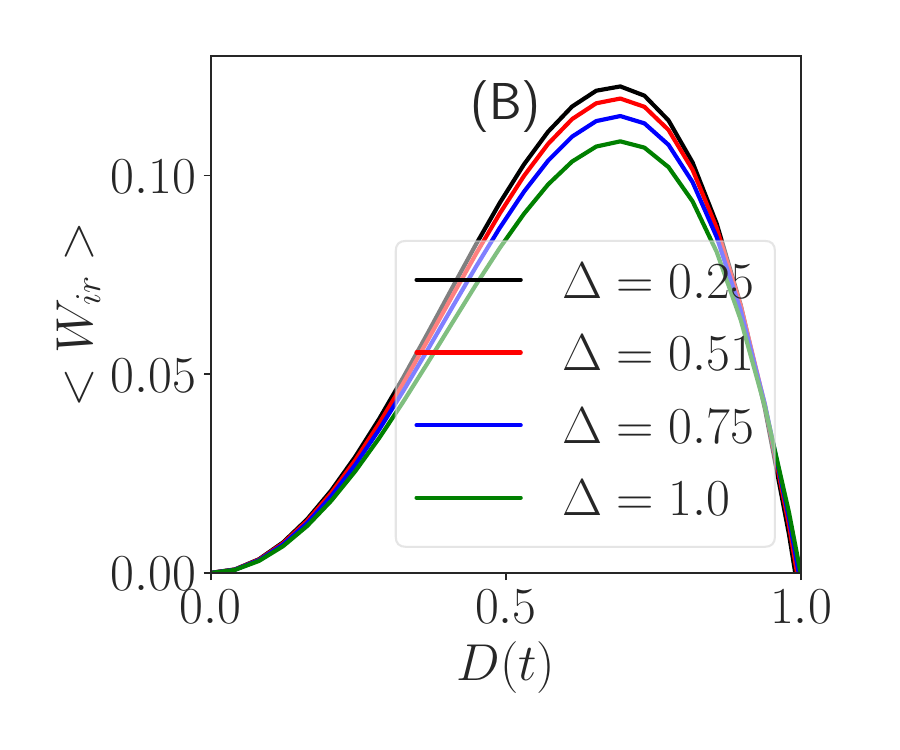} 
    \caption{(A) Irreversible work {\it vs.} $D$ for different time rates with which $D$ is varied. We steer the magnetic field such that we go from the ferromagnetic phase of the system to the skyrmionic phase. Here we consider $\Delta=0.51J$ and $v_1<v_2<v_3<v_4$.  (B) Irreversible work v/s $D$ for different $\Delta$ and rate $v=v_1$. Inverse temperature, $\beta=0.5$ in both the cases }
    \label{fig:irreversible_work_vs_D_different_rates}
\end{figure}

The irreversible work $\langle W_{ir}\rangle$ is influenced by the driving protocol and tends to increase for faster driving rates. In our setup, we keep the starting and ending points of the plasmonic electric field fixed at $D_0$ and $D_1$, respectively, and we achieve the variation of $D$ by implementing a linear quench in the Hamiltonian with different rates $v$. The functional dependence of $D$ on time during the quench is assumed to be as follows:
\begin{equation}
    D_t = D_0 + (D_1 - D_0)vt
    \label{dmi_time_dependance}
\end{equation}

As we see from Fig.~\ref{fig:irreversible_work_vs_D_different_rates} (A) and Fig.~\ref{fig:irreversible_work_vs_D_different_rates} (B), the irreversible work depends on the anisotropy parameter $\Delta$ and the quenching rate $v$. It also depends on the temperature $\beta$. In particular cases, irreversible work ($\langle W_{ir} \rangle$) at the end of the entire stroke cycle discussed earlier is almost zero (black and red lines) at feasible parameter regimes. This is endowed by the robustness and topological protection of the skyrmion state. 
Due to the zero $\langle W_{ir} \rangle$ Eq:~\ref{work done} gets simplified, and when calculating the efficiency (Work output / Heat input) of the skyrmion quantum Otto cycle, we exploit the formula:

\begin{eqnarray}
\eta=\frac{\Delta F_2+\Delta F_4}{Q_{in}}.
\end{eqnarray}

Here $\Delta F_2=Z(T_H,D_1)/Z(T_H,D_0)$, $\Delta F_4=Z(T_L,D_0)/Z(T_L,D_1)$, 
$T_L$, $T_H$ are temperatures of the cold and hot heat baths, $Q_{in}=\sum\limits_n \mathcal{E}_n(D_0)\left(P_n(T_H,D_0)-P_n(T_L,D_0)\right)$ is the heat pumped into the working substance and $P_n(T_H,D_0)=Z^{-1}(T_H,D_0)e^{-\mathcal{E}_n(D_0)/T_H}$,  $P_n(T_L,D_0)=Z^{-1}(T_L,D_0)e^{-\mathcal{E}_n(D_0)/T_L}$ are level populations.

\begin{figure}[!ht]
\centering
\includegraphics[scale=.5,trim={0.0cm 0.0cm 0.0cm 0.0cm},clip]{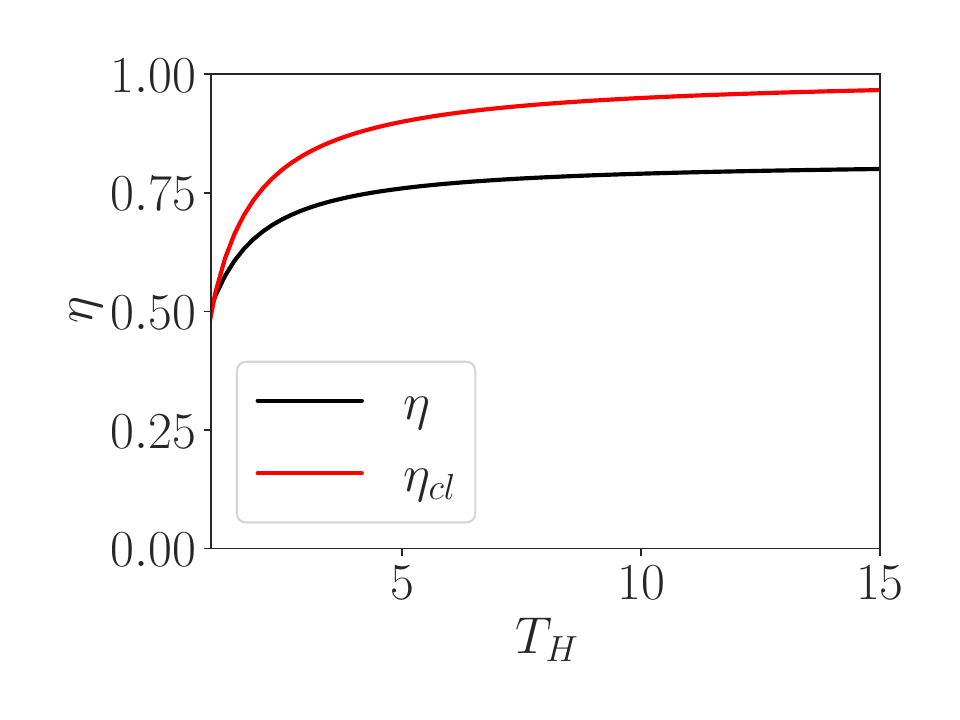}
\caption{The efficiency of the plasmonic skyrmion quantum heat engine, $\eta$ as a function of the hot bath's temperature $T_H>T_L$. $D_0=0J$ $D_1=2J$, $T_L=0.5$, $\Delta=0.25J$. $\eta_{cl}=1-T_L/T_H$ is the classical limit of the efficiency.}
\label{fig:efficiency}
\end{figure}
From Fig.~\ref{fig:efficiency}, we see that the efficiency of the cycle reaches 80 percent for the specified set of parameters and is quite high and well within the classical limit {\it i.e,}  $\eta<1-T_L/T_H$ limit. We can tune the efficiency by controlling factors like interaction parameter and $T_L$. The choice of parameters was based on two key factors. First is the skyrmion's stability, given by $Q$, and the second is the entropy production, governed by parameters leading to zero $\langle W_{ir}\rangle$ (see supplementary material for more detail). In conclusion the model we explored shows excellent potential to be considered for a heat engine.

{\bf Summary:}Quantum heat engines are one of the intriguing problems handled in quantum thermodynamics. The output from a nano scale quantum heat engines is affected by quantum friction caused by the inter level transitions. Conventionally this problem is dealt with introducing shortcuts to adiabaticity. Quantum skyrmion working substances can overcome quantum fricion thanks to the innate properties of its eigen states. The zero irreversible work done imply a practivally zero quantum friction loss. The heat engine model we propossed shows efficiencies as high as $80\%$.

\section{Acknowledgment}
We acknowledge National Supercomputing Mission (NSM) for providing
computing resources of ‘PARAM Shivay’ at Indian Institute of
Technology (BHU), Varanasi, which is implemented by C-DAC and
supported by the Ministry of Electronics and Information Technology
(MeitY) and Department of Science and Technology (DST), Government of
India. A.E. acknowledges funding by Fonds zur Förderung der
Wissenschaftlichen Forschung (FWF) Grant No. I 5384. This project has
received funding from the European Union’s Horizon 2020 research and
innovation programme under Grant Agreement No. 766566 (ASPIN) and from
the Deutsche Forschungsgemeinschaft (DFG, German Research Foundation) - project No. 403505322, Priority Programme (SPP) 2137. 
The work of M.I.K. was supported by the European Research Council (ERC)
under the European Union’s Horizon 2020 research and innovation program, Grant Agreement No. 854843-FASTCORR.
SKM acknowledges Science and Engineering Research Board, Department of Science and Technology, India, for support under Core Research Grant CRG/2021/007095.
\bibliography{2nems}

\end{document}


\renewcommand{\vec}[1]{\mathbf{#1}}
\newcommand{\ii}{\mathrm{i}}
\def\ya#1{{\color{orange}{#1}}}

%
%
%
%
%

\title{Supplimentary material to `Plasmonic skyrmion quantum thermodynamics'}

\date{\today}
\maketitle
\section{the irreversible work}\label{the irreversible work}
Our main statement is that due to the topological protection, irreversible work is small. 
Let us look at the quantities entering the expression of irreversible work:
\begin{eqnarray}
&&\langle W_{irv}\rangle=\frac{1}{\beta}S\left(\hat\varrho_t||\hat\varrho_t^{eq}\right),\\
&&S\left(\hat\varrho_t||\hat\varrho_t^{eq}\right)=\sum\limits_np_n^0\ln p_n^0-\sum\limits_{n,m}p^0_n p_{m,n}^\tau\ln p^\tau_n,\\
&&\hat\varrho_t^{eq}=e^{-\beta\hat E(t)}/\text{Tr}\left[e^{-\beta\hat H(t)}\right],\\
&&\hat\varrho_t=\hat U^{-1}(t)\hat\varrho(0)\hat U(t),\,\,\hat U(t)=e^{-i/\hbar\int\limits_0^t \hat H(\tau)d\tau},\\
&&p^\tau_{m,n}=|\bra{E_n(\tau)}\hat U(\tau)\ket{E_m(0)}|^2,\,\,Z(\beta,E(0))=\sum\limits_n e^{-\beta E_n(0)},\\
&&p_n^0=e^{-\beta E_n(0)}Z(\beta,E(0)),\,\,p^\tau_n=\sum\limits_m p^\tau_{m,n}p^0_m.
\end{eqnarray}

\begin{figure}[ht]
\centering
\begin{tabular}{c c}
\includegraphics[width=0.5\textwidth] {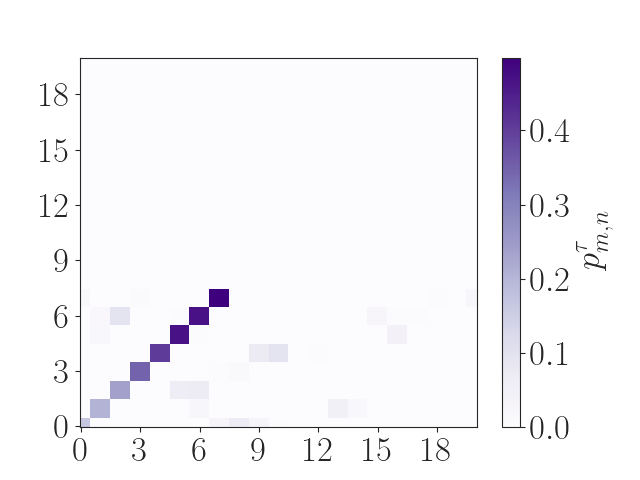} & \includegraphics[width=0.5\textwidth]{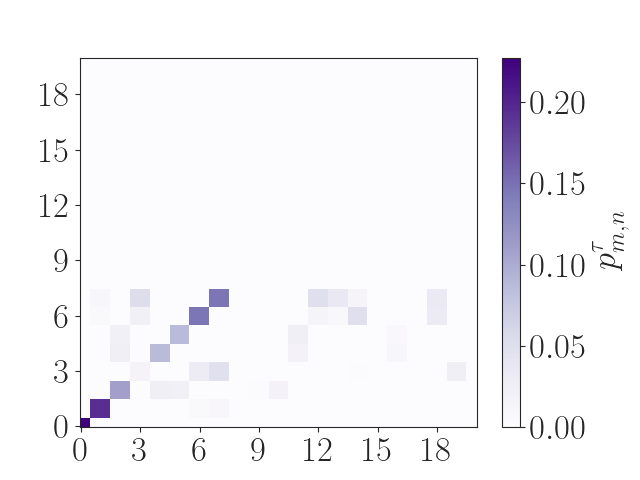} \\
(a) & (b)
\end{tabular}
\centering
\label{fig:transition_rate}
\caption{(a) An instance of $p^\tau_{m,n}$ is plotted for first few eigen states for $\Delta = 0.25J t$, $D_0=0.0J$ and $D_1=0.4J$. We see that only $m=n$ terms have significant contribution. Here the initial state is a skyrmion state and it is evolved within the parameter range where the system remains in skyrmionic ground state. (b) Here we plot the same quantity for $\Delta=0.75$. At $\Delta=0.75$ the irreversible entropy tend to settle at a large value after a similar evolution process as in (a). Here we see that the diagonal terms have diminishing magnitude also off diagonal terms have significant contribution compared to (a).}

\end{figure}

We steer the DMI constant from $D_0$ at $t=0$ to $D_1$ at $t=\tau$ and calculate $p^\tau_{m,n}$. The irreversible work is small or zero if $p^\tau_{m,n}\approx \delta_{nm}$. The results of numerical calculations are shown in Fig.~\ref{fig:transition_rate}). 

\begin{figure}[!ht]
\centering
\includegraphics[scale=.7,trim={0.0cm 0.0cm 0.0cm 0.0cm},clip]{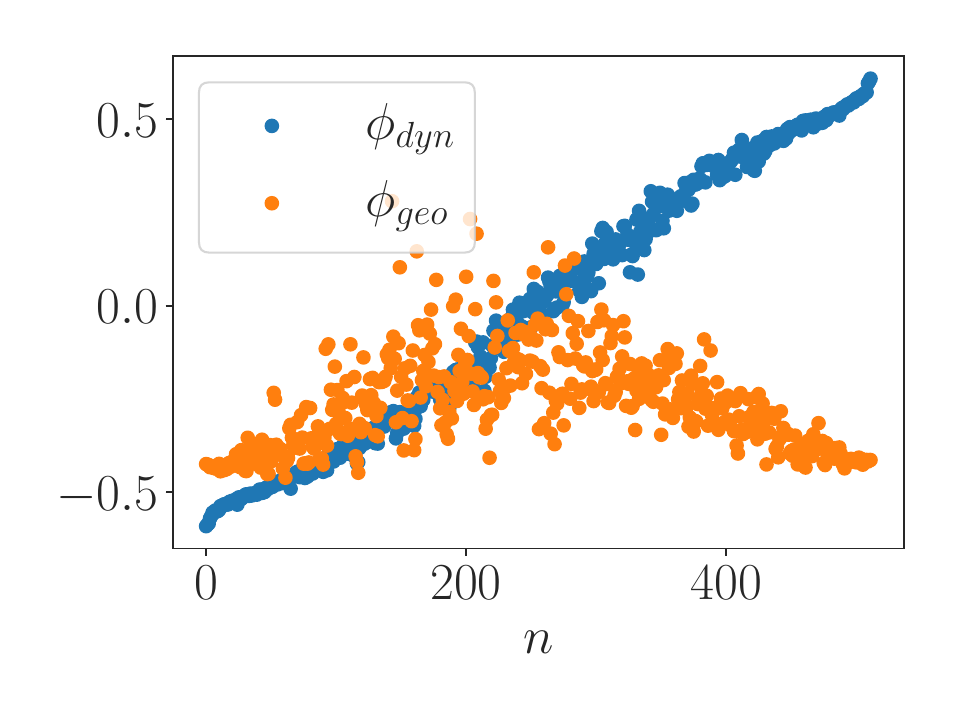}
\caption{Geometrical and dynamical phases obtained by wave functions during the adiabatic evolution in the skyrmion phase. Results are obtained for different $n$th states.}
\label{fig:geophase}
\end{figure}

The obtained result means that the propagated state $\ket{\psi(\tau)}=\hat U(\tau)\ket{E_n(0)}$ is different from the quantum skyrmion state 
$\ket{E_n(\tau)}$ only by geometric and dynamical phases, meaning that:

\begin{eqnarray}\label{phases}
\ket{\psi(\tau)}=\hat U(\tau)\ket{E_n(0)}=\exp\left[-\frac{i}{\hbar}\int\limits_0^\tau dt E_n(t)-\int\limits_0^\tau dt\langle E_n(t)\vert\partial_t E_n(t)\rangle\right]\ket{E_n(\tau)}.
\end{eqnarray}

Validity of Eq.(\ref{phases}) we checked by numerical calculations and observed that the propagated states differ only by the geometrical and the dynamical phases from the initial state. Results are plotted in Fig.\ref{fig:geophase}.